# Twist angle effects on the dynamic response of in-plane-switching liquid crystal displays

Yubao Sun[a)], Hongmei Ma, and Zhidong Zhang

*Department of Applied Physics, Hebei University of Technology, Tianjin, 300130, P. R. China*

Xinyu Zhu and Shin-Tson Wu

*College of Optics and Photonics, University of Central Florida, Orlando, Florida 32816*

**Abstract:**

Twist angle effect on the response time of in-plane-switching liquid crystal displays are analyzed. We propose a device configuration whose top and bottom boundary liquid crystal layers are symmetric to each other with respect to the electric field direction. The analytical results of this device configuration indicate that the response time is improved at least 4X faster than that of a conventional in-plane-switching twisted-nematic mode and normal in-plane- switching mode.

In-plane-switching (IPS) mode[1-3] has been widely used for large-screen transmissive liquid crystal displays (LCDs) because of its wide viewing angle, weak color shift, and high contrast ratio. However, the response time of the IPS mode is relatively slow as compared to other display modes because of its smaller twist elastic constant.[4] In the IPS mode, the effect of rubbing angle on LC director's response time has been analyzed in detail.[5-7] However, the twist angle effect has not yet been analyzed due to the difficulty in obtaining analytical solutions.

In this letter, we present two in-plane-switching twisted-nematic (TN) configurations and derive the corresponding LC director's dynamic response by solving the Erickson-Leslie equation under small angle approximation. The twist angle effects is analyzed quantitatively in the range of 0°~180°.

Figures 1(a) and 1(b) show the director configurations of two different IPS-TN modes under study. In both figures, $\Phi$ represents the total initial twist angle. In Fig. 1(a) the LC director on the bottom substrate is perpendicular to the electric field direction. However, in Fig. 1(b) both top and bottom boundary layers are symmetric with respect to the electric field direction. From Fig. 1(b), we find $\Phi=\pi-2\alpha$, where $\alpha$ is the initial LC alignment direction (i.e., rubbing angle) with respect to the vertical reference. Figure 1(a) is known as the IPS-TN mode when the total twist angle is $\Phi \leq \pi/2$.[8,9] On the other hand, if the twist angle is between $\pi/2$ and $\pi$, it is called IPS-STN,[10] as Fig. 1(b) depicts.

When the backflow and inertial effects are ignored, the dynamics of LC director rotation is described by the following Erickson-Leslie equation: [4,11]

$$\gamma_1 \frac{\partial \phi}{\partial t} = K_{22} \frac{\partial^2 \phi}{\partial z^2} + \varepsilon_o |\Delta\varepsilon| E^2 \sin\phi \cos\phi, \qquad (1)$$

where $\gamma_1$ is the rotational viscosity, $K_{22}$ is the twist elastic constant, $\Delta\varepsilon$ is the dielectric anisotropy, $E$ is the electric field strength, and $\phi$ is the LC director rotation angle. Here z axis is along the substrate normal direction and the LC cell gap is $d$. In our analysis, we assume the strong surface anchoring so that the bottom and top boundary layers are fixed even under applied voltage. For the IPS-TN cell shown in Fig. 1(a), the middle layer reaches the maximum twist angle deformation $\phi_m$ in the voltage-on state. Similar condition occurs at the quarter and three-quarter layers in the IPS-STN cell. Using these initial conditions, in the steady state we can approximately express the electric field induced twist angle $\phi$ deformation along the z axis as [12]

$$\phi(z) = \phi_1(z) + \phi_m \sin(\pi z/d), \tag{2a}$$

and

$$\phi(z) = \phi_2(z) + \phi_m \sin(2\pi z/d) \tag{2b}$$

for IPS-TN and IPS-STN, respectively. Here, $\phi_1(z) = \Phi z/d$ and $\phi_2(z) = \alpha + \Phi z/d$ are the twist profiles under zero electric field for Figs. 1(a) and 1(b), respectively.

First, let us consider the relaxation process. We assume the electric field is removed instantaneously at time t=0. During the relaxation period from the activated state to the initial state, the LC director deformation can be expressed by the following exponential forms:

$$\phi(z,t) = \phi_1(z) + \phi_m \sin(\pi z/d) \exp(-t/\tau), \tag{3a}$$

and

$$\phi(z,t) = \phi_2(z) + \phi_m \sin(2\pi z/d) \exp(-t/\tau), \tag{3b}$$

for IPS-TN and IPS-STN cells, respectively. Substituting Eqs. (3a) and (3b) into Eq. (1), the relaxation time can be solved easily:

$$\tau_{off} = \gamma_1 d^2 / (\pi^2 K_{22}), \tag{4a}$$

for the IPS-TN mode shown in Fig. 1(a), and

$$\tau_{off} = \gamma_1 d^2 / (4\pi^2 K_{22}), \tag{4b}$$

for the IPS-STN mode shown in Fig. 1(b).

Equations (4a) and (4b) indicate that the relaxation time of both twist configurations is governed by the cell gap $d$ and the LC visco-elastic coefficient ($\gamma_1/K_{22}$). Clearly, we find that the IPS-STN cell has 4X faster response time than the conventional IPS-TN mode. The LC directors at the middle plane of the IPS-STN mode, whose direction is parallel to the electric field direction, remain unchanged at the voltage-on state. Thus the IPS-STN is equivalent to two identical IPS-TN cells, each of which has one half of the LC layer thickness of the original cell. As compared to the IPS-TN cell, the IPS-STN cell should have 4X faster relaxation time provided that both configurations have the same cell gap $d$.

Next, we analyze the rise time. The LC director deformation during rise period can be described as: [12]

$$\phi(z,t) = \phi_1(z) + \phi_m \sin(\pi z/d)[1 - \exp(-2t/\tau)], \tag{5a}$$

for the IPS-TN cell and

$$\phi(z,t) = \phi_2(z) + \phi_m \sin(2\pi z/d)[1 - \exp(-2t/\tau)], \tag{5b}$$

for the IPS-STN cell. If $x_1 = \phi_m \sin(\pi z/d)$ and $x_2 = \phi_m \sin(2\pi z/d)$, then with the small angle approximation approach we have $x_i \ll 1$, $\cos(2x_i) \approx 1$, and $\sin(2x_i) \approx 2x_i$, where $i=1$ and 2. Substituting Eqs. (5a) and (5b) into Eq. (1), we obtain the rise time for the IPS-TN and IPS-STN cells as:

$$\tau_1 = 2\gamma_1 / \{\exp(2)\varepsilon_0 \Delta \varepsilon E^2 X_1 - \pi^2 K_{22}[\exp(2)-1]/d^2\}, \tag{6a}$$

and

$$\tau_2 = 2\gamma_1 \big/ \{\exp(2)\varepsilon_0\Delta\varepsilon E^2 X_2 - 4\pi^2 K_{22}[\exp(2)-1]/d^2\}, \tag{6b}$$

respectively, where $X_i = \int_0^d \{\sin[2\phi_i(z)] + 2x_i[1-\exp(2)]\cos[2\phi_i(z)]\}dz \big/ \int_0^d x_i dz$, i=1 and 2.

Under small angle approximation, Eqs. (6a) and (6b) can be, respectively, rewritten as:

$$\tau_1 = \gamma_1 \big/ \left[\varepsilon_0\Delta\varepsilon E^2/(2\phi_{m1}) - \pi^2 K_{22}/d^2\right], \tag{7a}$$

and

$$\tau_2 = \gamma_1 \big/ \left[\varepsilon_0\Delta\varepsilon E^2/(2\phi_{m2}) - 4\pi^2 K_{22}/d^2\right]. \tag{7b}$$

For a given weak electric field strength, $\phi_{mi}$(i=1,2) can be obtained from Eq. (1) and has an approximative relation $\phi_{m1} \approx 4\phi_{m2}$ for the two configurations with same twist angle ($\Phi$). From Eq.(7), the rise time of IPS-STN configuration is about 4X faster than that of IPS-TN configuration while the two configurations are driven by the same and low voltage.

Equations (4a), (4b), (6a), and (6b) are the general formulae for calculating the LC director's decay and rise time, assuming the applied voltage is not too high. To compare the switching time for different twist angles, the required electric field to obtain the same maximum twist angle deformation can be calculated from Eq.(1) and expressed as:

$$E_1^2 = \frac{K_{22}\pi^2}{d^2\varepsilon_0\Delta\varepsilon} \frac{\int_0^d x_1 dz}{\int_0^d \sin[\phi_1(z)+x_1]\cos[\phi_1(z)+x_1]dz}, \tag{8a}$$

for IPS-TN, and

$$E_2^2 = \frac{4K_{22}\pi^2}{d^2\varepsilon_0\Delta\varepsilon} \frac{\int_0^d x_2 dz}{\int_0^d \sin[\phi_2(z)+x_2]\cos[\phi_2(z)+x_2]dz}, \tag{8b}$$

for IPS-STN, respectively.

Using Eqs. 4(a), 4(b), 6(a), 6(b), 8(a), and 8(b), we can calculate the response time for the IPS-TN and for the IPS-STN cells. However, if the twist angle in Fig. 1(a) is between 90° and 180°, we should treat it as two cascaded TN cells with the virtual division line parallel to the electric field direction so that we can still calculate its response time. The first TN cell has an effective twist angle of 90° and an effective cell gap $d_{eff,1}=d\pi/2\Phi$, while the second TN cell has an effective twist angle $\Phi-90°$ and an effective cell gap $d_{eff,2}=d(1-\pi/2\Phi)$. For the case of $\pi/2<\Phi<\pi$, we find $d_{eff,1}>d/2$ and $d_{eff,2}<d/2$, which means the first TN cell has a larger effective cell gap than the second TN cell. Since the response time is proportional to the square of cell gap, under such a circumstance, the device response time is determined by the slowest one, which is the first cell because $d_{eff,1}>d_{eff,2}$. As a result, its response time can still be calculated by using Eqs. (4a), (6a), and (8a) as long as we substitute $d_{eff,1}=d\pi/2\Phi$ for $d$ as the twist angle in Fig. 1(a) is between 90° and 180°. In our calculation, the following LC parameters are used: $K_{22}=7$ $pN$, $\Delta\varepsilon=7.8$, d=3 $\mu m$, and $\gamma_1=0.1$ Pa.s.

To ensure that the small angle approximation holds, let us assume the maximum twist angle deformation is 0.05 *radian* from the initial state. The required driving electric field strength (I for Fig. 1(a) and II for Fig. 1(b)) with different twist angle is plotted in Fig. 2. The corresponding rise time and relaxation time for these two configurations are plotted in Figs. 3(a) and 3(b), respectively. From Fig. 3(a), we find that the rise time decreases as the total twist angle increases for the IPS-TN cell, but increases for the IPS-STN cell. More importantly, IPS-STN has a much shorter rise time than IPS-TN. The reasons are twofold. First, in order to have the same maximum twist angle deformation, IPS-STN cell needs a higher driving voltage than IPS-TN, as shown in Fig. 2. A larger driving electric field leads to a larger electric torque to the liquid crystal layer which makes the LC directors move faster.

Second, in the IPS-STN cell, the middle layer LC directors are not reoriented by the external electric field; therefore, we can assume that an invisible boundary layer exists in the middle. Consequently, the equivalent cell gap is only one half of the original one, resulting in a much faster dynamic response. From Fig. 3(b), we find that the relaxation time remain a constant (~3.25 ms) for the IPS-STN cell, as indicated in Eq. 4(b). For the IPS-TN cell, as we can see from curve I in Fig. 3(b), the relaxation time keeps unchanged as the twist angle increases from 0° to 90° and then quickly decreases as the twist angle further increases from 90° to 180° because of the variety of the effective cell gap, which is discussed above. When the twist angle equals to 180°, the effective cell gap $d_{eff,1}=d/2$, which is identical to the IPS-STN cell shown in Fig. 1(b), the same response time can be seen from Fig. 3. Overall speaking, the device configuration shown in Fig. 1(b) exhibits a faster response time than that of Fig. 1(a) when the total twist angle is less than 180°.

In conclusion, we have analyzed the effect of twist angle on the response time of in-plane switching mode LCDs. The response time depends strongly on the twist angle and the initial director configuration. Both rise time and relaxation time of IPS-STN configuration are faster than those of IPS-TN configuration due to the unactivated middle LC layer in the IPS-STN configuration. Our analysis results will help to improve the dynamic response performance of IPS mode LCDs.

This research was supported by the Key Construction Project of Hebei Provincial University and Natural Science Foundation of Hebei Province (No. A2006000675 and No. 103002), P. R. China.

**Figure Captions:**

FIG. 1 Two different LC director profiles in the LC cell for (a) conventional IPS-TN mode with LC director on the bottom substrate perpendicular to the electric field and (b) symmetric IPS-STN mode with LC director of the middle layer parallel to electric field.

FIG. 2 The required driving electric field for $\phi_m=0.05$ in the two configurations with different twist angle. Curve I is for the configuration of Fig.1(a), and curve II is for the configuration of Fig.1(b).

FIG. 3 The rise time (a) and relaxation time (b) for the two configurations with different twist angle. Curve I is for the configuration of Fig.1(a), and curve II is for the configuration of Fig.1(b).

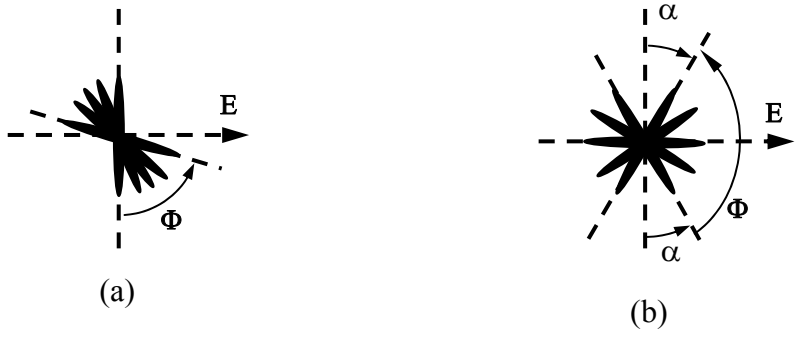

Fig. 1
Sun et al

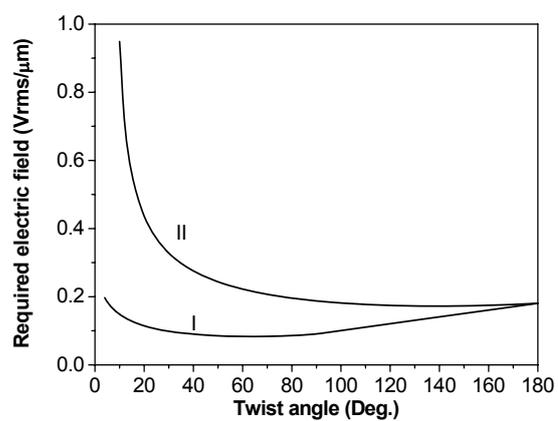

Fig. 2
Sun et al

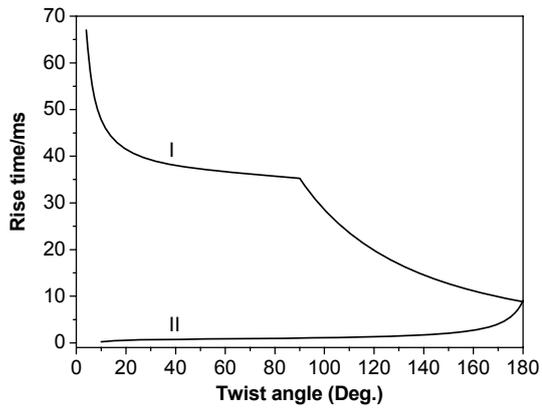

(a)

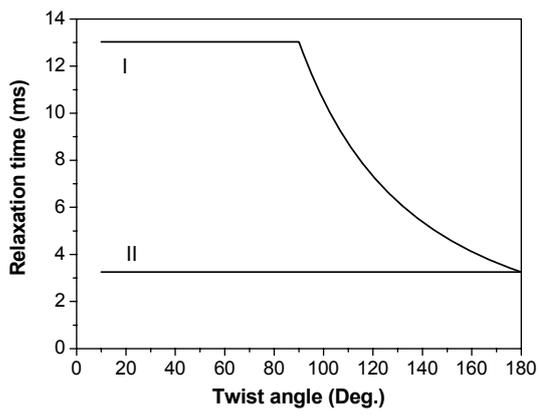

(b)

Fig. 3
Sun et al